\documentstyle[preprint,pra,aps]{revtex}

\begin{document}

\draft

\title{Macroscopic Quantum Tunneling of a Bose Condensate}

\author{H.T.C. Stoof}
\address{University of Utrecht, Institute for Theoretical
         Physics, Princetonplein 5, \\
         P.O. Box 80.006, 3508 TA  Utrecht, The Netherlands}

\maketitle

\begin{abstract}
We study, by means of a variational method, the stability of a condensate in a
magnetically trapped atomic Bose gas with a negative scattering length and find
that the condensate is unstable in general. However, for temperatures
sufficiently close to the critical temperature the condensate turns out to be
metastable. For that case we determine in the usual WKB approximation the decay
rate of the condensate due to macroscopic quantum fluctuations. When
appropriate, we also calculate the decay rate due to thermal fluctuations. An
important feature of our approach is that (nonsingular) phase fluctuations of
the condensate are taken into account exactly.
\end{abstract}


\section{INTRODUCTION}
\label{int}
The observation of Bose-Einstein condensation in dilute atomic $^{87}$Rb
\cite{JILA}, $^7$Li \cite{rice}, and $^{23}$Na \cite{MIT} vapors last year, has
created a great deal of excitement in the atomic physics community. Although it
was speculated upon already for some time, the actual observation of the
condensation phenomenon in these alkali gases nevertheless came as somewhat of
a surprise, because in the spring of 1995 experiments with a trapped atomic
hydrogen gas still held the record in achieving the necessary conditions for
Bose-Einstein condensation \cite{H1,H2}. The main reasons for this unexpected
turn of events appear to be that alkali atoms are much more easy to detect and
that experiments with alkali vapors can be performed at room temperature in
contrast with the cryogenic environment that is required for experiments with
an atomic hydrogen gas.

The report of Bose-Einstein condensation in atomic $^7$Li was also surprising
for another, more fundamental, reason. It had namely been established
experimentally \cite{randy1} that the effective interaction between two $^7$Li
atoms is attractive, or more precisely that the s-wave scattering length $a$ is
negative. For a homogeneous gas this implies that Bose-Einstein condensation
cannot take place in the mechanically (meta)stable region of the phase diagram
and is preempted by a first-order gas-liquid or gas-solid transition
\cite{henk}. Although first Hulet \cite{randy2} and subsequently also Ruprecht
{\it et al.} \cite{keith} had suggested that this conclusion holds in an
inhomogeneous situation only if the number of atoms is sufficiently large, the
number of atoms used in the experiment indeed appeared to be too large by about
two orders of magnitude.

At present, this unsatisfactory state of affairs still exists, i.e.\ experiment
claims to observe a condensate in a gas with a negative scattering length
whereas theory seems to predict that this should not be possible. In an attempt
to bridge at least part of the gap between theory and experiment, we study in
Sec.~\ref{tun2} for zero and subsequently also for nonzero temperatures, the
stability of a condensate of $^7$Li atoms in an isotropic harmonic oscillator
potential. In the process of this analysis, we show that there exists an
interesting analogy between the quantum dynamics of the condensate and the
quantum mechanics of a particle in an unstable potential. Therefore, to make
the paper more selfcontained and to bring out this analogy most clearly, we
present first in Sec.~\ref{tun1} a brief summary of how the WKB approximation
to the tunneling rate of a particle is derived by path-integral methods. We end
in Sec.~\ref{dis} with some conclusions.

\section{TUNNELING OF A PARTICLE}
\label{tun1}
The tunneling rate $\Gamma_0$ of a particle with mass $m^*$ out of a metastable
minimum of a potential $V(q)$ can be calculated by means of the relation
\begin{equation}
\Gamma_0 = -\frac{2}{\hbar} {\rm Im}(E_0)
         = \lim_{T \downarrow 0}~
                      \frac{2k_BT}{\hbar} {\rm Im}({\rm ln}Z)~,
\end{equation}
where ${\rm Im}(E_0)$ is the imaginary part of the (analytically continued)
groundstate energy in the metastable minimum of the potential, and $Z =
Tr[e^{-\beta H}]$ is the partition function with $\beta = 1/k_BT$ and $H(p,q) =
p^2/2m^* + V(q)$ the usual Hamilton operator for the particle. This is a
convenient starting point for our discussion, because the partition function
can in the usual way be represented as a functional integral over the functions
$q(\tau)$ and $p(\tau)$, i.e.\
\begin{equation}
\label{Z}
Z = \int d[q] \int d[p] ~
    \exp \left\{ - \frac{1}{\hbar}
                     \int_{-\hbar\beta/2}^{\hbar\beta/2} d\tau
           \left( -i p \frac{d q}{d\tau} + H(p,q) \right)
         \right\}
\end{equation}
with the periodic boundary condition
$q(-\hbar\beta/2) = q(\hbar\beta/2)$ on the coordinate but no restrictions on
the momentum. For the hamiltonian of interest the integral over the momentum
$p(\tau)$ is just a gaussian that can be easily carried out. As a result the
partition function is in this case also equal to a path integral
$\int d[q] \exp \left\{- S[q]/\hbar \right\}$ over all periodic paths $q(\tau)$
and with a (Euclidian) action given by
\begin{equation}
\label{SE}
S[q] = \int_{-\hbar\beta/2}^{\hbar\beta/2} d\tau
           \left( \frac{1}{2} m^*
                        \left( \frac{d q}{d\tau}
                        \right)^2 + V(q)
           \right)~.
\end{equation}

Our task is therefore to evaluate this path integral. In general this cannot be
done exactly, but one can obtain a good (semiclassical) approximation by noting
that the dominant contributions to the path integral are from paths that
minimize the action $S[q]$. Such paths are solutions to the Euler-Lagrange
equation
\begin{equation}
\label{EL}
m^* \frac{d^2 q}{d\tau^2} = - \frac{d(-V(q))}{dq} ~,
\end{equation}
which has the same form as the classical equation of motion for a particle with
mass $m^*$ in a potential $-V(q)$. A periodic solution to this equation is
therefore $q(\tau) = q_0$, where $q_0$ is the position of the metastable
minimum and obeys $dV(q_0)/dq = 0$. Writing $q(\tau) = q_0 + q'(\tau)$ and
expanding the action up to quadratic order in the fluctuations $q'(\tau)$, we
obtain first of all that
\begin{equation}
Z \simeq  e^{-\beta V(q_0)}
      \int d[q'] \exp
        \left\{ -\frac{1}{\hbar}
           \int_{-\hbar\beta/2}^{\hbar\beta/2} d\tau
           \left( \frac{1}{2} m^*
                        \left( \frac{d q'}{d\tau}
                        \right)^2
              + \frac{1}{2} \frac{d^2 V(q_0)}{dq^2} q'^2
           \right)
        \right\}~.
\end{equation}
Introducing $m^* \omega_0^2 \equiv d^2 V(q_0)/dq^2$, we notice that the path
integral in the right-hand side is just equal to the partition function for a
harmonic oscillator and therefore that
\begin{eqnarray}
\int d[q'] \exp
        \left\{ -\frac{1}{2}
           \int_{-\hbar\beta/2}^{\hbar\beta/2} d\tau~
           q' \left( -\frac{m^*}{\hbar} \frac{d^2}{d\tau^2}
              + \frac{m^*\omega_0^2}{\hbar} \right) q'
        \right\}    \hspace*{2.0in} \nonumber \\
= {\cal N} \left\{ {\rm det} \left[ \frac{m^*}{2\pi\hbar}
                    \left( - \frac{d^2}{d\tau^2} + \omega_0^2
                    \right)
                    \right] \right\}^{-1/2}
= \exp \left\{ -\beta \frac{\hbar\omega_0}{2}
          - {\rm ln} \left( 1 - e^{-\beta\hbar\omega_0} \right)
       \right\}~,
\end{eqnarray}
denoting the usual normalization factor due to the measure in the path integral
by ${\cal N}$. Taking now the limit $T \downarrow 0$ (or $\beta \rightarrow
\infty$) we thus find in this approximation for the groundstate energy $E_0
\simeq V(q_0) + \hbar\omega_0/2$ and for the tunneling rate $\Gamma_0 = 0$.
This is clearly a reasonable first-order result that is due to the fact that we
can in first instance always approximate the full potential $V(q)$ by the
harmonic oscillator potential
$V(q_0) + m^*\omega_0^2(q-q_0)^2/2$.

To obtain a nonzero value for the tunneling rate we must realize that there is
another periodic solution to Eq.~(\ref{EL}) that has an action which (in the
zero temperature limit) is only slightly different from $S[q_0]$ and therefore
also gives an important contribution to the partition function. Using the
classical analogy, this so-called `bounce' solution $q_b(\tau)$ has the
property that the particle spends a very long time around $q_0$ but in a
relatively short time oscillates once in the potential minimum of $-V(q)$, i.e.
it bounces from $q_0$ to $q_1$ and back to $q_0$, where $q_1$ obeys $V(q_1) =
V(q_0)$. In particular, for $\tau \rightarrow \pm \infty$ it behaves as
$q_b(\tau) \sim q_0 \mp (v_0/\omega_0) e^{-\omega_0|\tau|}$ with $v_0$
determined by the details of the potential.

Proceeding as before by writing $q(\tau) = q_b(\tau) + q'(\tau)$ and expanding
the action up to quadratic order in $q'(\tau)$, we now find that near zero
temperature the partition function equals
\begin{equation}
Z \simeq e^{\beta(V(q_0) + \hbar\omega_0/2)}
    \left( 1 + \left[ \frac{{\rm det}(-d^2/d\tau^2 + \omega_0^2)}
                           {{\rm det}(-d^2/d\tau^2 +
                                               \omega_b(\tau)^2)}
               \right]^{1/2} e^{-(S[q_b]-S[q_0])/\hbar}
    \right)~,
\end{equation}
introducing the quantity $\omega_b(\tau)$ by means of
$m^* \omega_b(\tau)^2 \equiv d^2 V(q_b(\tau))/dq^2$. Adding in a similar manner
also the contributions from paths with an arbitrary number of bounces, we find
that the series exponentiates and hence that
\begin{equation}
\label{E}
E_0 \simeq  V(q_0) + \frac{\hbar\omega_0}{2}
        - ~\lim_{T \downarrow 0}~ k_BT
               \left[ \frac{{\rm det}(-d^2/d\tau^2 + \omega_0^2)}
                           {{\rm det}(-d^2/d\tau^2 +
                                               \omega_b(\tau)^2)}
               \right]^{1/2} e^{-(S[q_b]-S[q_0])/\hbar}~.
\end{equation}
We expect the third term in the right-hand side to represent the tunneling rate
out of the metastable minimum and therefore to be purely imaginary. This
expectation is indeed correct, because the operator $-d^2/d\tau^2 +
\omega_b(\tau)^2$ turns out to have a negative eigenvalue. This is most easily
understood from the fact that Eq.~(\ref{EL}) shows that $d q_b(\tau)/d\tau$ is
an eigenfunction of this operator with an eigenvalue equal to zero. Since this
eigenfunction has one node, we know from our experience with the Schr\"odinger
equation that there must be an eigenfunction without nodes that has a lower,
and therefore, negative eigenvalue.

However, the presence of an eigenvalue equal to zero appears to give an
infinite result for the ratio of determinants in Eq.~(\ref{E}). Fortunately,
this is due to an improper treatment of the zero mode in the calculation of the
path integral over the fluctuations $q'(\tau)$. Since
$q_b(\tau-\tau_0) = q_b(\tau)
                    - \tau_0 dq_b(\tau)/d\tau + O(\tau_0^2)$,
we note that this zero mode is just associated with a translation of the
`bounce' solution $q_b(\tau)$ and hence that the square root of the ratio of
determinants must be proportional to the total time interval
$\hbar\beta = \hbar/k_BT$. A detailed and beautiful analysis by Duru {\it et
al.} actually shows that
\begin{equation}
\lim_{T \downarrow 0}~ k_BT
  \left[ \frac{{\rm det}(-d^2/d\tau^2 + \omega_0^2)}
              {{\rm det}(-d^2/d\tau^2 + \omega_b(\tau)^2)}
               \right]^{1/2}
  = \frac{i}{2} \sqrt{ \frac{m^* v_0^2 \hbar\omega_0}{\pi} }~.
\end{equation}
Combining this with Eq.~(\ref{E}) we finally arrive at
\begin{equation}
\Gamma_0 = - \frac{2}{\hbar} {\rm Im}(E_0)
         = \sqrt{ \frac{m^* \omega_0 v_0^2}{\pi\hbar} }~
             e^{-(S[q_b]-S[q_0])/\hbar}~,
\end{equation}
where
\begin{equation}
\frac{S[q_b]-S[q_0]}{\hbar} =
  \frac{2}{\hbar} \int_{q_0}^{q_1} dq~
                              \sqrt{2m^*(V(q)-V(q_0))}
\end{equation}
is recognized as the usual WKB expression for the exponent of the tunneling
rate.

\section{TUNNELING OF A CONDENSATE}
\label{tun2}
We now turn to the problem of the stability of the condensate in a magnetically
trapped gas of $^7$Li atoms. We consider here only the experimentally relevant
case of a large number of particles $N \gg 1$ in a large trap. Quantitatively,
the latter means that $r_V/\ell \ll 1$, where $r_V$ is the range of the
interatomic interaction, $\ell=\sqrt{2\pi\hbar/m\omega}$ is the spatial extent
of the one-particle ground state in a harmonic oscillator potential with level
spacing $\hbar\omega$, and $m$ is the mass of the $^7$Li atoms. (In the case of
$^7$Li the above condition also implies that $|a|/\ell \ll 1$.) Moreover, we
will always require that the density $n$ in the center in the trap is such that
the gas parameter $nr_V^3 \ll 1$. This basically leads to an upper bound on the
total number of particles that is always satisfied by the experiment of
interest \cite{rice}.

\subsection{The case ${\bf T = 0}$}
\label{T0}
At zero temperature and in an external trapping potential
$V^{ext}(\vec{x}) = m\omega^2\vec{x}^2/2$, the effective hamiltonian for the
condensate wavefunction
$\psi(\vec{x},t)$ and its canonical momentum
$\pi(\vec{x},t) = i\hbar \psi^*(\vec{x},t)$ is given by
\begin{equation}
H[\pi,\psi] = \int d\vec{x}~
   \psi^*
     \left( - \frac{\hbar^2}{2m} \nabla^2 + V^{ext}
            + \frac{T^{2B}(\vec{0},\vec{0};0)}{2}
                                         |\psi|^2
     \right)
   \psi
\end{equation}
since then the Hamilton equations exactly reproduce the nonlinear Schr\"odinger
equations for $\psi(\vec{x},t)$ and $\psi^*(\vec{x},t)$, respectively. For
example, we correctly find that
\begin{eqnarray}
\label{NLSE}
\frac{\partial}{\partial t} \psi(\vec{x},t)
 &\equiv& \frac{\delta}{\delta \pi(\vec{x},t)}
                                   H[\pi,\psi]    \nonumber \\
 &=& \frac{1}{i\hbar}
       \left( - \frac{\hbar^2}{2m} \nabla^2 + V^{ext}(\vec{x})
              + T^{2B}(\vec{0},\vec{0};0)
               |\psi(\vec{x},t)|^2
       \right) \psi(\vec{x},t)~,
\end{eqnarray}
with $T^{2B}(\vec{0},\vec{0};0) = 4\pi a \hbar^2/m$ the effective interaction
between the atoms. It is important to mention here that, due to infrared
divergences in the theory of the dilute Bose gas \cite{N}, this nonlinear
Schr\"odinger equation is only valid if the energy cut-off provided by the
trapping potential is sufficiently large. More precisely we must require that
$\hbar\omega > 4\pi n|a|\hbar^2/m$. Fortunately, this correponds precisely to
the conditions under which the condensate is metastable as we will find out
shortly.

After this brief discussion of the condensate wavefunction as a classical
field, we can now turn to the quantum fluctuations of the condensate. As in
Sec.~\ref{tun1}, we must then consider the partition function $Z=Tr[e^{-\beta
H}]$ at zero temperature. This function can be written as the functional
integral
\begin{equation}
\label{Z'}
Z = \int d[\psi^*] d[\psi]~
      \exp \left\{ - \frac{1}{\hbar} \int d\tau
             \left( \int d\vec{x}~ \psi^*
                      \hbar \frac{\partial}{\partial\tau} \psi
                  + H[i\hbar\psi^*,\psi]
             \right)
           \right\}
\end{equation}
over the periodic fields $\psi(\vec{x},\tau)$ and $\psi^*(\vec{x},t)$, which is
the direct analog of Eq.~(\ref{Z}). To proceed as in the case of the quantum
mechanics of a particle, we should now integrate out the momentum field. This
is, however, not helpful for our purposes because the hamiltonian does not have
a momentum independent part. Consequently, it seems that a stability analysis
of the condensate cannot be performed in the same way as in
Sec.~\ref{tun1}.

The way out of this dilemma is found by noting that the instability of the
condensate is, just as in the homogeneous case, a result of density
fluctuations that lead to a lower energy because of the effectively attractive
interaction between the $^7$Li atoms. It is therefore advantageous to perform a
canonical transformation by means of the relation
$\psi = \sqrt{\rho}e^{i\chi}$ and to use the density field $\rho(\vec{x},\tau)$
and the phase field $\chi(\vec{x},\tau)$ to calculate the partition function.
Indeed, a simple substitution together with the periodicity of
$\rho(\vec{x},\tau)$ shows that the partition function is equal to the
functional integral
$\int d[\rho] \int d[\chi] \exp\{-S[\rho,\chi]/\hbar\}$ with an action
\begin{equation}
\label{S}
S[\rho,\chi] = \int d\tau \int d\vec{x}~
    \left(   i\hbar \rho \frac{\partial \chi}{\partial\tau}
           + \frac{\hbar^2 \rho}{2m} (\nabla \chi)^2
           + \frac{\hbar^2}{8m\rho} (\nabla \rho)^2
           + V^{ext} \rho
           + \frac{T^{2B}(\vec{0},\vec{0};0)}{2} \rho^2
    \right)
\end{equation}
that is quadratic in the phase field $\chi(\vec{x},\tau)$. Hence, we can now
immediately integrate over this field.

There is an important point to be made about this integration, which reflects
the fact that if the original fields $\psi(\vec{x},\tau)$ and
$\psi^*(\vec{x},\tau)$ are periodic, the phase field $\chi(\vec{x},\tau)$ is
only periodic up to a multiple of $2\pi$. To calculate the partition function
correctly, we must therefore first integrate over all the fields
$\chi(\vec{x},\tau)$ obeying the boundary condition
$\chi(\vec{x},\infty) = \chi(\vec{x},-\infty) + 2\pi j$ and subsequently sum
over all possible integers $j$. Clearly, this change in boundary conditions
affects only the zero-momentum part of $\chi(\vec{x},\tau)$. As a result we
have to consider the sum
\begin{eqnarray}
\sum_{j} \int d[\chi_{\vec{0}}]~
   \exp \left\{-i \int d\tau~ N_0
                  \frac{d\chi_{\vec{0}}}{d\tau}
        \right\}                            \nonumber
\end{eqnarray}
first, where we made use of the fact that the total number of particles in the
condensate
$N_0(\tau) = \int d\vec{x} \rho(\vec{x},\tau)$ and the boundary condition
$\chi_{\vec{0}}(\infty) = \chi_{\vec{0}}(-\infty) + 2\pi j$ on the integration
is implicitly assumed for each term in the sum. Performing a partial
integration on the integral in the exponent, this sum becomes equal to
\begin{eqnarray}
\sum_j e^{2\pi i N_0 j}~ \delta \left[ \frac{dN_0}{d\tau}
                                \right]              \nonumber
\end{eqnarray}
because the integration over $\chi_{\vec{0}}(\tau)$ then simply leads to the
constraint of a constant number of particles in the condensate, i.e.\
$N_0(\tau) = N_0$. At zero temperature and under the conditions that the
nonlinear Schr\"odinger equation is valid, this actually implies that
$N_0(\tau)$ is equal to the total number of particles $N$. In addition, the sum
$\sum_j e^{2\pi i N_0 j}$ equals $\sum_j \delta(N_0 - j)$ and thus requires
that $N_0$ is an integer. In this manner we see explicitly that the integration
over the zero-momentum part of $\rho(\vec{x},\tau)$ is effectively only a sum
over the total number of particles and, most important for our purposes, that
the partition function at a constant number of particles is given by the
functional integral $\int d[\rho] \int d[\chi] \exp\{-S[\rho,\chi]/\hbar\}$
over all nonzero momentum components of the density and phase fields.

The integration over the nonzero momentum part of $\chi(\vec{x},\tau)$ is
readily accomplished by introducing the Green's function for the phase
fluctuations $G(\vec{x},\vec{x}';\rho)$, that obeys
\begin{equation}
\frac{\hbar}{m} \left( (\nabla \rho) \cdot \nabla
      + \rho \nabla^2 \right) G(\vec{x},\vec{x}';\rho) =
                                 \delta(\vec{x}-\vec{x}')~.
\end{equation}
In terms of this Green's function the formal solution to the Euler-Lagrange
equation $\delta S[\rho,\chi]/\delta\chi = 0$ reads
\begin{equation}
\chi(\vec{x},\tau) = -i \int d\vec{x}'~
         G(\vec{x},\vec{x}';\rho)
             \frac{\partial \rho(\vec{x}',\tau)}{\partial \tau}~,
\end{equation}
which after a substitution in Eq.~(\ref{S}) gives the following effective
action for the density field
\begin{eqnarray}
S[\rho] = \int d\tau \int d\vec{x} \int d\vec{x}'~
    \left(-\frac{\hbar}{2} \frac{\partial \rho(\vec{x},\tau)}
                                 {\partial \tau}
             G(\vec{x},\vec{x}';\rho)
                            \frac{\partial \rho(\vec{x}',\tau)}
                                 {\partial \tau}
    \right)                       \hspace*{0.5in}  \nonumber \\
        + \int d\tau \int d\vec{x}~
             \left(
               \frac{\hbar^2}{8m\rho} (\nabla \rho)^2
               + V^{ext} \rho
               + \frac{T^{2B}(\vec{0},\vec{0};0)}{2} \rho^2
             \right)~.
\end{eqnarray}
This is the desired analog of Eq.~(\ref{SE}) that we will now use to determine
the stability of the condensate at zero temperature.

Our first task is to see whether a metastable condensate is possible at all in
a trap. This question can be answered by considering only the `potential
energy'
\begin{equation}
V[\rho] = \int d\vec{x}~
             \left(
               \frac{\hbar^2}{8m\rho} (\nabla \rho)^2
               + V^{ext} \rho
               + \frac{T^{2B}(\vec{0},\vec{0};0)}{2} \rho^2
             \right)
\end{equation}
for time-independent density configurations. Because we are unable to
analytically consider all configurations, we proceed in a variational way
\cite{baym} and investigate here only the gaussian profiles
\begin{equation}
\rho(\vec{x};q) = N \left( \frac{1}{\pi q^2} \right)^{3/2}
                  \exp \left( -\frac{\vec{x}^2}{q^2} \right)~.
\end{equation}
The physical reason behind this choice is that we expect the shape of the
condensate, if it is metastable, to be close to the square of the one-particle
groundstate wavefunction of the trapping potential and, therefore, to be
reasonably accurately described by a gaussian. We will see shortly that this
expectation is indeed correct.

Substituting the above gaussian into the expression for the `potential energy',
we find that
\begin{equation}
\label{pot}
V[\rho] =
 N \left(
      \frac{3\hbar^2}{4m q^2} + \frac{3}{4} m \omega^2 q^2
       - \frac{N}{\sqrt{2\pi}}
              \frac{\hbar^2|a|}{mq^3}
   \right) \equiv N V(q)~.
\end{equation}
Hence, for $|a| = 0$ the potential $V(q)$ has an absolute minimum at $q =
\sqrt{\hbar/m\omega} = \ell/\sqrt{2\pi}$ and the `potential energy' at this
minimum is $3N\hbar \omega/2$. Clearly, this corresponds exactly to an ideal
Bose condensate. For $|a| \neq 0$ the `potential energy' is always unbounded
from below because $V(q) \rightarrow -\infty$ if $q \downarrow 0$. This implies
that the condensate is in general unstable and tends to collapse to the density
profile
$\lim_{q \downarrow 0} \rho(\vec{x};q) = N \delta(\vec{x})$. Of course, in a
realistic system the collapse to this density profile is ultimately prevented
by the hard core of the interatomic interaction as the density reaches a value
of the order of $O(1/r_V^3)$. Although our theory certainly breaks down at
these densities, the approximation $N \delta(\vec{x})$ for the final density
profile is nevertheless rather accurate for the lengthscales of interest, due
to the condition $r_V/\ell \ll 1$.

The most important feature of Eq.~(\ref{pot}) is, however, that if the
condition
\begin{equation}
\label{meta}
N \frac{|a|}{\ell} < \frac{2}{5^{5/4}} \simeq 0.27
\end{equation}
is fullfiled, the potential $V(q)$ has a metastable minimum. This result can
directly be compared with the work of Ruprecht {\it et al.}, who find by a
numerical integration of the nonlinear Schr\"odinger equation Eq.~(\ref{NLSE})
that a stable solution can only be obtained if $N |a|/\ell < 0.23$
\cite{keith}. Our variational calculation, therefore, gives an upper bound that
is only 16$\%$ too high. Apparently this is the amount of accuracy that one can
obtain for the condensate energy by considering only the density profiles
$\rho(\vec{x};q)$.

Having arrived at the conclusion that a metastable condensate is possible at
zero temperature if the number of particles in the gas is sufficiently small,
we now want to calculate the lifetime of the condensate due to macroscopic
quantum fluctuations. This involves also the evaluation of the `kinetic energy'
\begin{equation}
T[\rho] =
 \int d\vec{x} \int d\vec{x}'~
    \left(-\frac{\hbar}{2} \frac{\partial \rho(\vec{x},\tau)}
                                 {\partial \tau}
             G(\vec{x},\vec{x}';\rho)
                            \frac{\partial \rho(\vec{x}',\tau)}
                                 {\partial \tau}
    \right)~.
\end{equation}
Again an exact treatment of this part of the effective action is unfeasible.
However, our previous discussion of the `potential energy' suggests a
reasonably accurate approximation that amounts to the promotion of the
variational parameter $q$ to a real dynamical variable $q(\tau)$. Hence, we
assume that for a description of the tunneling process the most important
configurations of the condensate are given by
$\rho(\vec{x},\tau) = \rho(\vec{x};q(\tau))$. The reason for this assumption is
threefold. First, we have already seen that the profile $\rho(\vec{x};q)$ gives
an accurate description of the metastable minimum of the potential $V[\rho]$.
Second, one expects that the minimum energy barrier for the tunneling process
is associated with a rotationally symmetric configuration \cite{cole} that,
because of the diluteness of the gas, can be represented by a gaussian with the
same amount of accuracy as the metastable minimum. Third, the profile
$\rho(\vec{x};q)$ describes in the limit $q \downarrow 0$ also the completely
collapsed state of the system. In combination, the configurations
$\rho(\vec{x};q(\tau))$ thus seem to provide a reasonable interpolation between
the initial and final density profiles of the tunneling process.

Evaluating the `kinetic energy' for the variational ansatz $\rho(\vec{x},\tau)
= \rho(\vec{x};q(\tau))$ and adding
Eq.~(\ref{pot}) we easily find that the quantum dynamics of the condensate is
determined by the action
\begin{equation}
S[q] = N \int d\tau
           \left( \frac{1}{2} m^*
                        \left( \frac{d q}{d\tau}
                        \right)^2 + V(q)
           \right)~,
\end{equation}
which differs only by an overall factor $N$ from the zero temperature limit of
Eq.~(\ref{SE}). The effective mass $m^*$ is difficult to calculation exactly
due to the inhomogeneity of the system, but can be estimated by noting that the
dominant contributions to the `kinetic energy' come from the region near the
center of the trap. In this region the Green's function for the phase
fluctuations is well approximated by
\begin{equation}
G(\vec{x},\vec{x}';\rho) \simeq
  - \frac{m}{4\pi\hbar \rho(\vec{0};q)}
                 \frac{1}{|\vec{x}-\vec{x}'|}
\end{equation}
and we obtain
\begin{equation}
m^* \simeq \frac{m}{\pi^{5/2}}
  \int d\vec{y}\int d\vec{y}'~
     \left(\frac{3}{2} - \vec{y}^2\right) \exp{(-\vec{y}^2)}
                  \frac{1}{|\vec{y}-\vec{y}'|}
     \left(\frac{3}{2} - \vec{y}'^2\right) \exp{(-\vec{y}'^2)}
  \simeq 0.27 m~.
\end{equation}
Although the exact value of the effective mass is presumably somewhat larger
than this result, it is in any case of the order of the atomic mass $m$.

Applying now the final result of Sec.~\ref{tun1} (replacing only $\hbar$ by
$\hbar/N$) we immediately find that the decay rate of the condensate is equal
to
\begin{equation}
\label{rate}
\Gamma_0 = \sqrt{ \frac{Nm^* \omega_0 v_0^2}{\pi\hbar} }~
   \exp \left\{ - \frac{2N}{\hbar}
                    \int_{q_0}^{q_1} dq~
                              \sqrt{2m^*(V(q)-V(q_0))}
        \right\}
\end{equation}
and typically very small for a number of condensate particles
$N \gg 1$ but still sufficiently small to fulfill the condition of
metastability. Note that both requirements do not exclude each other, because
we are only considering traps for which $|a|/\ell \ll 1$.

\subsection{The case ${\bf T \neq 0}$}
We have seen that at zero temperature a metastable condensate is possible if
the number of $^7$Li atoms is less than
$N_m \equiv 0.23 \ell/|a| \simeq 1440$. (The scattering length is $-27 a_0$ and
the effective isotropic level splitting
$\hbar(\omega_x \omega_y \omega_z)^{1/3}$ for the trap is
$7 nK$ \cite{rice}.) In the Rice experiment, however, one reports the
observation of a condensate for temperatures as high as $400 nK$ and total
numbers of atoms as great as $2 \cdot 10^{5}$. Clearly, for these temperatures
the previous discussion is inadequate and needs to be modified before we can
draw any conclusions about a possible discrepancy between theory and
experiment.

The generalization to nonzero temperatures is essentially only useful in the
case that the number of particles $N \gg 100$, since then the critical
temperature $T_c$ is almost equal to the critical temperature
\begin{equation}
T_0 = \left( \frac{N}{\zeta(3)} \right)^{1/3}
             \frac{\hbar\omega}{k_B}
    \simeq 0.94 N^{1/3}~\frac{\hbar\omega}{k_B}
\end{equation}
of the ideal Bose gas and obeys $k_B T_c \gg \hbar\omega$ so that we are close
to the thermodynamic limit. As a result also the thermal wavelength
$\Lambda_c = \sqrt{2\pi\hbar^2/mk_BT_c} \ll \ell$.
To understand why this is important, we must realize that at nonzero
temperatures the nonlinear Schr\"odinger equation for the condensate
wavefunction is
\begin{equation}
\label{NLSET}
i\hbar \frac{\partial}{\partial t} \psi(\vec{x},t)
  = \left( - \frac{\hbar^2}{2m} \nabla^2 + V^{ext}(\vec{x})
           + T^{2B}(\vec{0},\vec{0};0)
             \left( 2n'(\vec{x},t) + |\psi(\vec{x},t)|^2 \right)
    \right) \psi(\vec{x},t)~,
\end{equation}
with $n'(\vec{x},t)$ the local density of atoms which are not in the condensate
\cite{tony}, and that the total number of noncondensed particles is of
$O((k_BT/\hbar\omega)^3)$ because $n'(\vec{0},t)$ is of $O(1/\Lambda^3)$ and
the size of the noncondensed cloud is of $O(\ell^2/\Lambda)$. Therefore, for
temperatures that do not obey $k_B T \gg \hbar\omega$, the term
$2n'(\vec{x},t)$ in Eq.~(\ref{NLSET}) is negligible compared to
$|\psi(\vec{x},t)|^2$. For temperatures $T \gg \hbar\omega/k_B$ this does not
have to be the case. However, under these conditions we can replace
$2n'(\vec{x},t)$ by a constant since the size of the noncondensed cloud is now
much larger than the size of the condensate. Performing then the same analysis
as in Sec.~\ref{T0}, we find that the potential $V(q)$ is just shifted by a
constant.

We thus conclude that for $N \gg 100$ a metastable condensate can exists as
long as $N_0 < N_m$, or sufficiently close to the critical temperature.
Moreover, the decay rate of the condensate can at all temperatures be
calculated by means of the effective action
\begin{equation}
\label{st}
S[q] = N_0 \int_{-\hbar\beta/2}^{\hbar\beta/2} d\tau~
           \left( \frac{1}{2} m^*
                        \left( \frac{d q}{d\tau}
                        \right)^2 + V(q)
           \right)~.
\end{equation}
Hence, for $T \ll \hbar\omega_0/k_B$ quantum fluctuations dominate and the
decay rate is given by Eq.~(\ref{rate}) with $N$ replaced by $N_0$. For $T \gg
\hbar\omega_0/k_B$, which is the relevant temperature interval if $N \gg N_m$,
the decay of the condensate is dominated by thermal (or classical) fluctuations
and the decay rate is proportional to the Boltzmann factor
$e^{-N_0(V(q_m)-V(q_0))/k_BT}$, where $q_m$ denotes the position at which
$V(q)$ has a maximum. The prefactor of the exponent is always a difficult
point, but can be estimated by noting that in the classical limit the action of
Eq.~(\ref{st}) implies a probability distribution
\begin{equation}
P(p,q) \simeq \frac{\beta\omega_0}{2\pi}
  \exp \left\{ -\beta
         \left( \frac{p^2}{2m^*N_0} + N_0(V(q)-V(q_0)) \right)
       \right\}
\end{equation}
for the momentum $p$ and the coordinate $q$ of a fictitious classical particle.
Using this probability distribution we then easily find from the average flux
over the energy barrier that
\begin{equation}
\Gamma_0 = \frac{\omega_0}{2\pi}
                \exp \left\{ - \frac{N_0}{k_BT} (V(q_m)-V(q_0))
                     \right\}~.
\end{equation}
Notice that this decay rate is only small if
$N_0(V(q_m)-V(q_0)) \ll k_BT$. Therefore, a long-lived metastable condensate is
only possible sufficiently close to the critical temperature, if the total
number of atoms obeys $N \ll (N_m)^3$.

\section{CONCLUSIONS}
\label{dis}
In view of the results obtained above it seems that we have arrived at an
explanation of the experiment with atomic $^7$Li, since we have shown that a
long-lived metastable condensate is possible if $1 \ll N_0 < N_m$.
Unfortunately, this is only partly  true because we have not yet considered the
mechanical stability of the noncondensed cloud. In particular, we have not
shown that the gas is stable (or metastable) at the critical temperature and
will, in contrast to what occurs in the homogeneous case \cite{henk}, not
immediately collapse to a dense liquid or solid phase. In our opinion, this is
at present the most important question which remains to be answered before we
can speak of a theoretical understanding of the Rice experiment.

Finally, we need to discuss the important observation by Dalfovo and Stringari
that a condensate with a vortex line through the center of the trap has a value
of $N_m$ that is significantly larger than $0.23 \ell/|a|$, due to the fact
that the local density of the gas must vanish at the point where the phase
singularity occurs \cite{stringari}. In principle, we can also consider the
decay of a condensate with a vortex in the framework of the theory presented in
Sec.~\ref{tun2}. We should then first put
$\chi(\vec{x},\tau) = \chi_v(\vec{x},\tau) +
                                        \chi'(\vec{x},\tau)$,
where $\chi_v(\vec{x},\tau)$ is the phase configuration of a vortex, before we
integrate over the nonsingular phase fluctuations $\chi'(\vec{x},\tau)$. The
main difficulty, however, will be to find an accurate solution of the coupled
dynamical equations for the density profile and the position of the vortex.
Although such a calculation would certainly be interesting in its own right, it
clearly does not invalidate the main conclusion of our paper that the most
important unresolved issue is the stability of the gas without a condensate.

After completion of the work presented in Sec.~\ref{T0} of this paper, we
received a preprint by Kagan, Shlyapnikov, and Walraven \cite{kagan} in which
these authors also consider the stability of an inhomogeneous condensate at
$T=0$, and arrive at qualitatively the same results. However, their calculation
of the decay rate is completely different from ours and, in particular, does
not take phase fluctuations of the condensate into account. As a result, it
appears not to correspond to a WKB treatment of the tunneling process.

\section*{ACKNOWLEDGMENTS}
I would like to thank Randy Hulet for various enlightening communications and
for providing me with a preprint of Ref.\ \cite{keith} that inspired the above.
I also thank Keith Burnett and Peter Ruprecht for stimulating discussions on
this work at the now historic BEC workshop in Mont Ste Odile. Finally, I am
grateful to Michel Bijlsma and Marianne Houbiers for a careful reading of the
manuscript.

\end{document}